# Patient-specific simulation of stent-graft deployment within an abdominal aortic aneurysm

**David PERRIN\*, Pierre BADEL\*, Stéphane AVRIL\*, Jean-Noël ALBERTINI\*\*, Laurent ORGEAS\*\*\*, Christian GEINDREAU\*\*\*, Aurélien DUMENIL\*\*\*\*, Cemil GOKSU\*\*\*\***

\* Ecole Nationale Supérieure des Mines de Saint-Etienne, CIS-EMSE, CNRS:UMR5307, LGF, F-42023 Saint Etienne, France, {perrin ; badel ; avril}@emse.fr

\*\* CHU Hôpital Nord Saint-Etienne, Department of CardioVascular Surgery, Saint-Etienne F-42055, France, j.noel.albertini@chu-st-etienne.fr

\*\*\* CNRS / Université de Grenoble (Grenoble-INP / UJF), Laboratoire Sols-Solides-Structures-Risques (3SR Lab), BP 53, 38 041 Grenoble cedex 9, France,{laurent.orgeas ; christian.geindreau}@3sr-grenoble.fr

\*\*\*\* Therenva, 35000 Rennes, France, {aurelien.dumenil ; cemil.goksu}@therenva.fr

## SUMMARY

In this study, finite element analysis is used to simulate the surgical deployment procedure of a bifurcated stent-graft on a real patient's arterial geometry. The stent-graft is modeled using realistic constitutive properties for both the stent and most importantly for the graft. The arterial geometry is obtained from pre-operative imaging exam. The obtained results are in good agreement with the post-operative imaging data. As the whole computational time was reduced to less than 2 hours, this study constitutes an essential step towards predictive planning simulations of aneurysmal endovascular surgery.

# 1 INTRODUCTION

Endovascular aneurysm repair (EVAR) is a widely and increasingly used technique to treat abdominal aortic aneurysms (AAAs) owing to several assets related to the minimally invasive approach. However, to date, the durability of stent-graft (SG) treatments remains the principal issue of EVAR. Secondary interventions after 5 years are required in up to 22% of cases due to endoleaks, stenosis or thrombosis of the SG or failure of the SG's constituents; also, in tortuous AAAs, a lack of SG flexibility was associated with the above-mentioned complications [1]. These facts clearly emphasize the need to predict, as soon as the surgical planning, the mechanical response of SGs in a given geometry. In pursuit of this goal, FE analyses can be used for predicting the deployment of stent-grafts, hence the initial success or failure of the procedure.

In the literature, the expansion of bare stents has been the object of intensive research and many numerical studies [2]. However, the deployment of SG – widely used in the treatment of aortic aneurysms – has been the object of very few studies [3,4]. And yet, the presence of the textile component onto which the stents are sutured is a key aspect which drastically influences the behavior of the SG and requires proper modeling. Provided that the mechanical properties of each constituent are cautiously calibrated, numerical simulations of the deformation of SGs [5] has the potential to provide important indications to aid a surgeon: guiding the surgeon in choosing the right dimensions for the stent-graft, especially lengths, in patient-specific aneurysm models and anticipating (i) potential short-term complications such as stent-graft kinks or inadequate apposition onto the arterial wall, (ii) potential long-term complications such as stent rupture or fabric tear. In short, introducing the appropriate SG models into simulations of the deployment would be very useful in surgical planning.

The goal of the present study was, thus, to develop a model able to simulate the positioning of a bifurcated SG in a given patient-specific AAA, under static conditions. Significant effort was made to build a clinically-relevant and very cost-efficient simulation model within the perspective of using it as a tool to help clinicians in surgical planning.

# 2 METHODS

**Geometry and properties.** The geometry of the pathological aorta, including the abdominal aorta and the iliac arteries, was obtained from the pre-operative CT scan of the considered patient, followed by segmentation of the lumen and 3-node shell meshing of the wall using the surgery-oriented EndoSize® software (Therenva, France). The stent-graft which was implanted in this patient was modeled as a multi-component structure including the textile graft onto which the metallic stents are sewn. The graft was modeled with 4-node shell elements and the stents with 2-node beam elements tied to the textile.

Since only the *in vivo* tangent behavior is considered, the artery's constitutive model was assumed as orthotropic elastic with Young's moduli of 1 and 0.5 MPa in circumferential and longitudinal directions. One important aspect for the simulation was the proper characterization and modeling of the in-plane and bending behaviors of the textile due to their significant impact on the simulation results (kinks and fabric wrinkles). A realistic orthotropic constitutive model was set up based on in-house experimental characterization [5]. The simulation strategy (presented below) restraining the stents to their elastic response, a linear elastic model was assigned to these components (Young's modulus E = 40 GPa, and Poisson's ratio $\upsilon$ = 0.46).

**Simulation strategy.** To make the simulation as efficient as possible within the perspective of being clinically tractable and relevant, the main focus of this work was dedicated to reducing the computational cost to obtain a relevant final deployed state of the SG. Hence, the numerous actual steps of the deployment procedure were simplified in two steps. In the first step, the SG is initially surrounded by a "virtual" shell containing the SG to be deployed, and contact between the shell and the SG is activated (see Fig. 1a). Then mesh morphing of this shell onto the actual arterial geometry is achieved by prescribing displacements to all the nodes of the shell (Fig. 1b). Stresses in the SG result from this step as the SG diameter is oversized of about 15% with regard to the proximal neck of the aorta. Then, in the second step, the shell has initially the shape of the original artery before SG implantation. After assigning the elastic properties of the artery to the shell and prescribing proper boundary conditions, the mechanical equilibrium between the SG and the artery is computed, hence the final geometry after SG implantation *in vivo* (Fig. 1c).

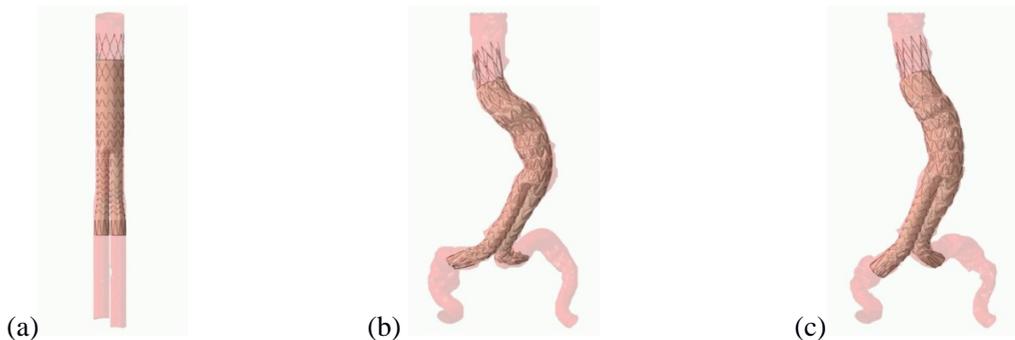

(a)          (b)          (c)
**Fig. 1**. (a) initial configuration of the model; (b) result of morphing onto the arterial geometry (step 1); (c) result of mechanical equilibrium (step 2).

## 3  RESULTS AND DISCUSSION

Figure 1 presents the initial geometry, the result of step 1 (morphing onto the patient's arterial geometry) and those of step 2 (mechanical equilibrium). Wrinkles are reproduced in a realistic manner. Figure 2 shows a comparison with post-operative geometry obtained from CT-scan images. The qualitative agreement is very encouraging. Some discrepancies can be noticed in iliac portions, which is likely due to boundary conditions in this region where the internal iliac arteries may locally constrain the artery (not considered yet). In addition, part of these differences may be due to the length of the iliac SG limbs and the overlap between the limbs and the main body of the SG which are not known.

Validation of the approach on a cohort of real cases is on-going. The results reported constitute the first step towards predictive planning simulations of aneurysmal endovascular surgery.

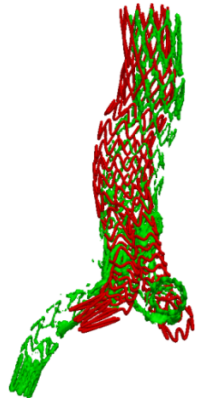

**Fig. 2**. Qualitative comparison of the computed SG geometry (in red) to the real SG geometry (in green) obtained from post-operative CT-scan.

## Acknowledgements

This work is a part of a thesis granted by the Region Rhône-Alpes.